\documentclass[twoside,slac_one]{revtex4}
\usepackage{graphicx}
\usepackage{fancyhdr}
\usepackage{amsmath} % American Mathematics Society standards
\usepackage{bm}% bold math
\usepackage{amsxtra}
\usepackage{amssymb}
\usepackage{amsthm}
\usepackage{latexsym}
\usepackage{lscape}

\begin{document}

\title{Search for \boldmath$WH\rightarrow\ell \nu b\bar{b}$ Final States at the Tevatron}


\author{D. P. Brown\\
~\\
On Behalf of the CDF and D0 Collaborations\\
~\\}
\affiliation{CNRS - LPNHE, Universities Paris VI and VII, 4 place Jussieu, Tour 33 - Rez de Chaussee, \\ 75252 Paris Cedex 05, France}

\begin{abstract}
Latest results are presented in the search for low mass standard model Higgs production 
in association with a $W$ boson, based on large luminosity data samples
collected at the CDF and D0 Experiments at the Fermilab Tevatron $p\bar{p}$ collider. The 
selection of event samples containing an isolated lepton, an imbalance in transverse
energy in the events, and either one or two reconstructed jets consistent with having 
evolved from a $b$-quark, provides statistically independent data samples to search for 
$q\bar{q}\rightarrow WH$ candidates. Expected and observed upper limits are derived for 
the product of the $WH$ production cross section and branching ratios and are reported
in units of the standard model prediction. The observed (expected) upper limits for a 
Higgs mass $M_{H}=115 ~\rm GeV$ are factors 2.65 (2.6) and 4.6 (3.5) above the 
standard model prediction for the CDF and D0 searches, respectively.
\end{abstract}

\maketitle

\thispagestyle{fancy}


%%%%%%%%%%%%%%%%%%%%%%%%%%%%%%%%%%
\section{Introduction}

In the Standard Model (SM) of particle physics the explanation for the 
finite masses of the weakly interacting $W$ and $Z$ bosons is the process 
of electroweak symmetry breaking yielding a single Higgs particle state. 
The finite masses of fermions are then accounted for via their Yukawa couplings 
to the Higgs field. At Tevatron $p\bar{p}$ collider energies, searches for SM Higgs 
production using $\ell \nu b\bar{b}$ final states are expected to be among the most 
sensitive to SM Higgs production, most notably in the Higgs mass range $100 < m_{H} < 135 ~\rm GeV$. 
The requirement of a $W$ Boson reconstructed in association with jets serves 
to reduce experimental backgrounds from QCD jet production processes, and improves 
sensitivity to signal.

The searches presented here are based on large luminosity data samples collected at the 
CDF and D0 experiments at the Tevatron $p\bar{p}$ collider. \cite{cdf,d0} 
Candidate $W\rightarrow e \nu $ and $W\rightarrow \mu \nu $ events are selected by requiring a 
single isolated lepton together with an associated imbalance in transverse energy in the events. 
The Higgs decay mode $H\rightarrow b\bar{b}$ is used since it has the largest expected branching 
fraction in the studied mass regime. Statistically independent (orthogonal) data samples are selected 
via the application of $b$-tagging. Multivariate techniques are then applied to suppress remaining search 
backgrounds in each sample. Finally, upper limits are derived for the product of the $WH$ production 
cross section and branching ratios and reported in units of the SM prediction.

Direct searches for the process $e^{+}e^{-}\rightarrow ZH$ at CERN $e^{+}e^{-}$ LEP collider 
experiments already constrain the minimum SM Higgs mass to $m_{H}> 114.4~\rm GeV$ at 95\% 
confidence level. In addition, a fit to precision electroweak measurements of the top-quark and W boson masses, 
from both Tevatron $p\bar{p}$ and CERN $e^{+}e^{-}$ Collider experiments,  infer an upper limit of 
$m_{H} < 161 ~\rm GeV$ at 95\% CL \cite{lep,fit} and more recently searches at the CERN LHC $pp$ 
collider experiments yield preliminary exclusions at larger Higgs masses. \cite{atlas,cms}

\section{Event Selections}

\begin{figure*}[ht]
\centering
\includegraphics[width=67mm]{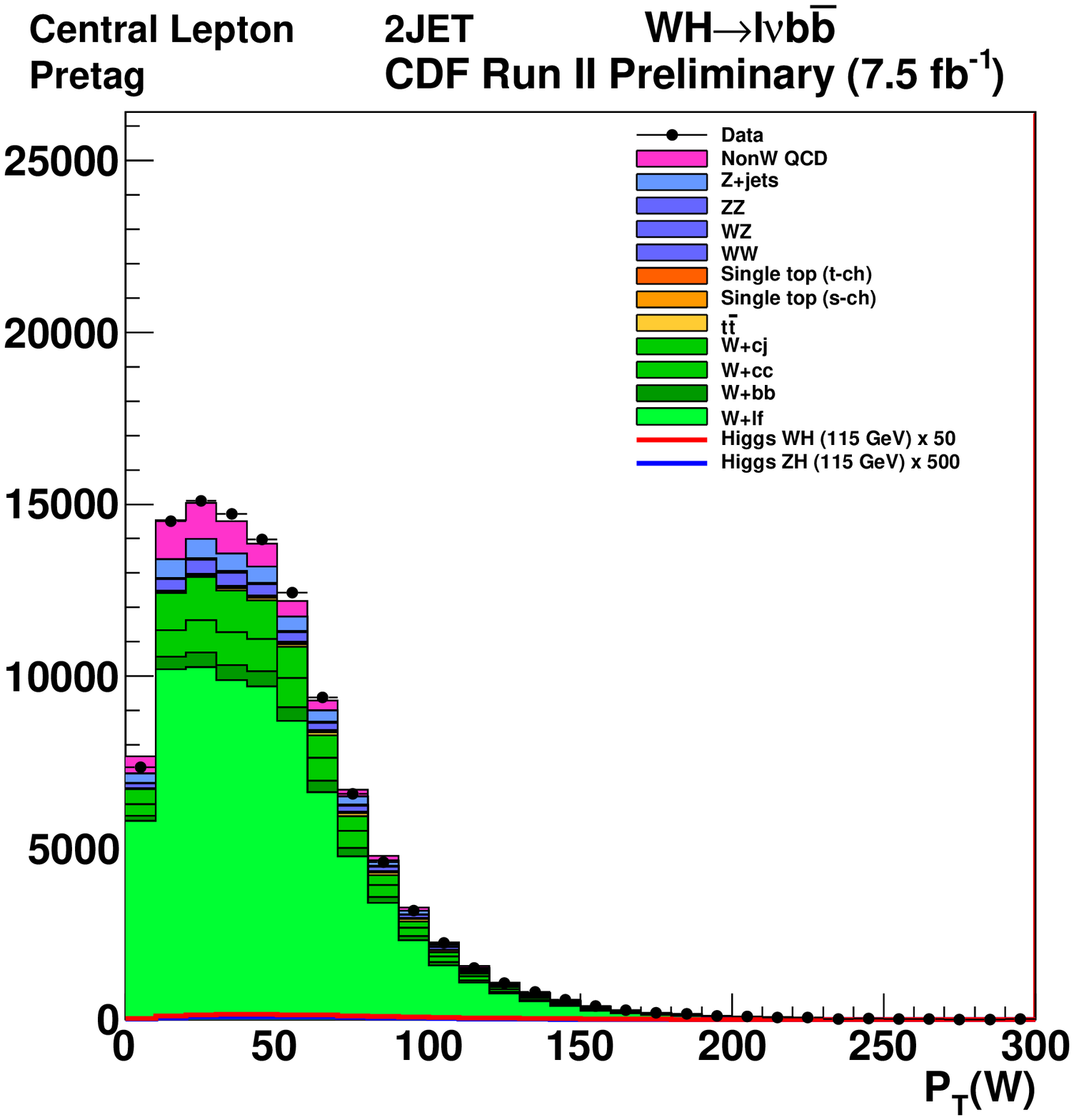}
\includegraphics[width=93mm]{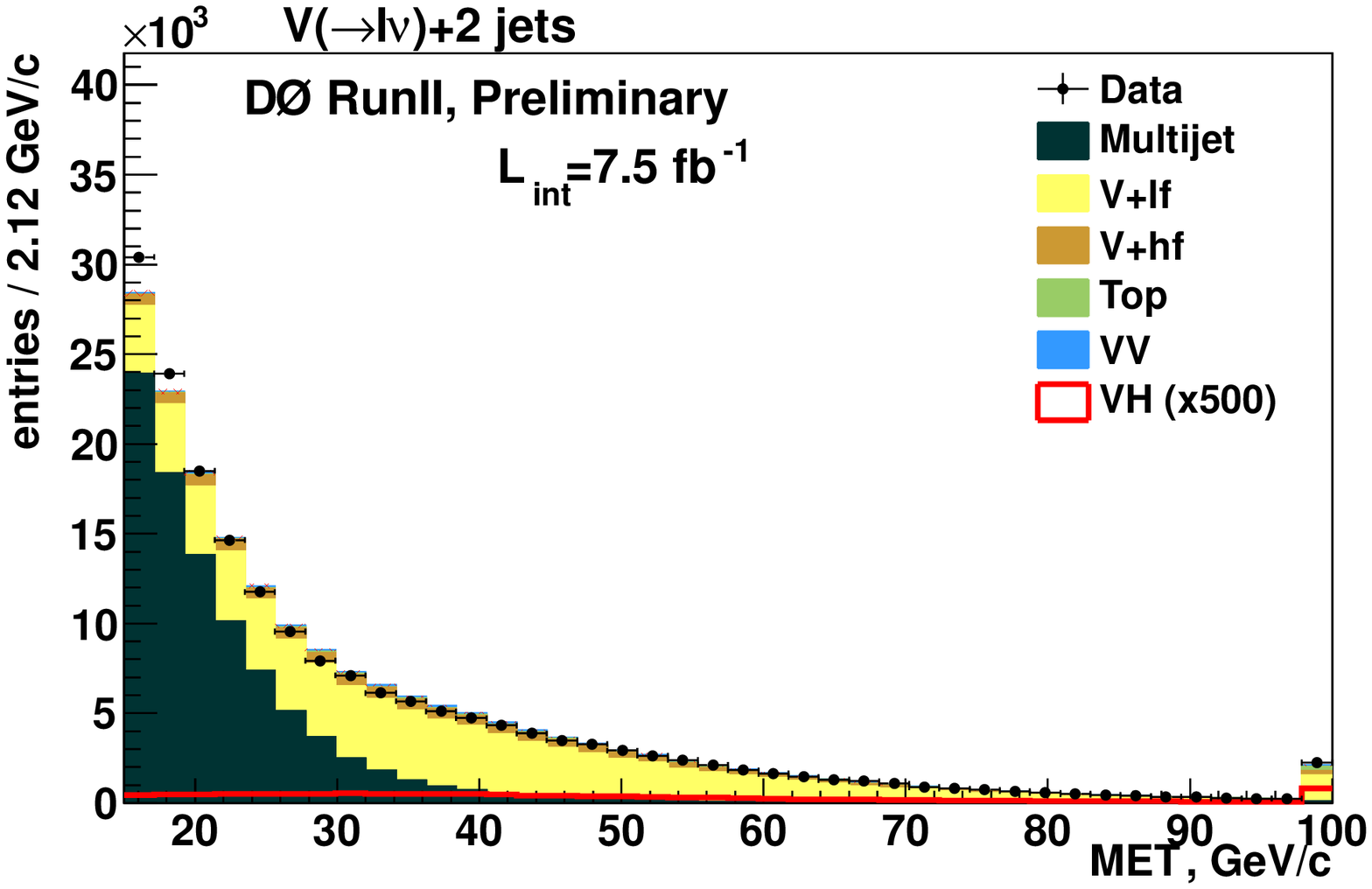}
\includegraphics[width=92mm]{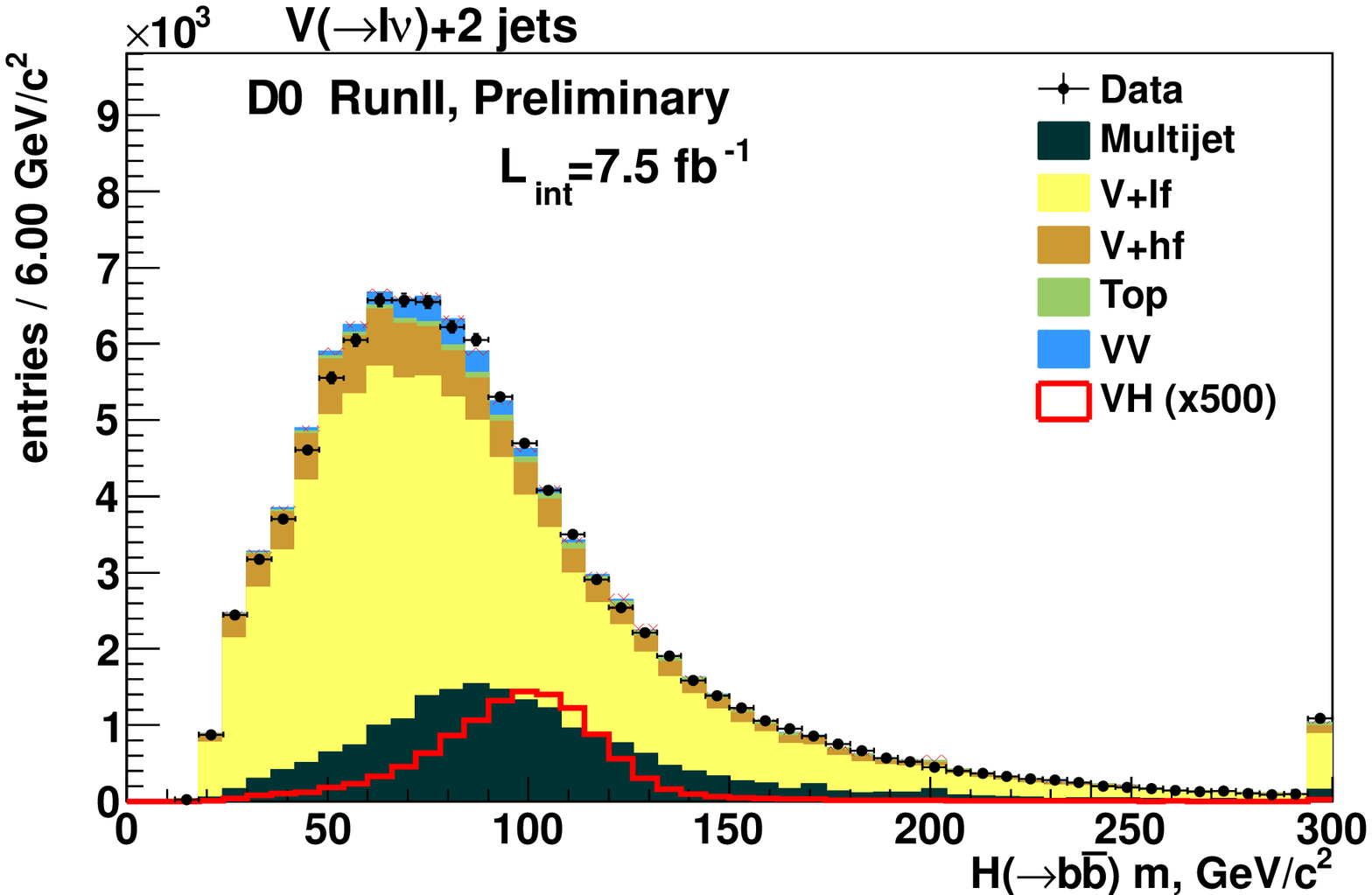}
\includegraphics[width=73mm]{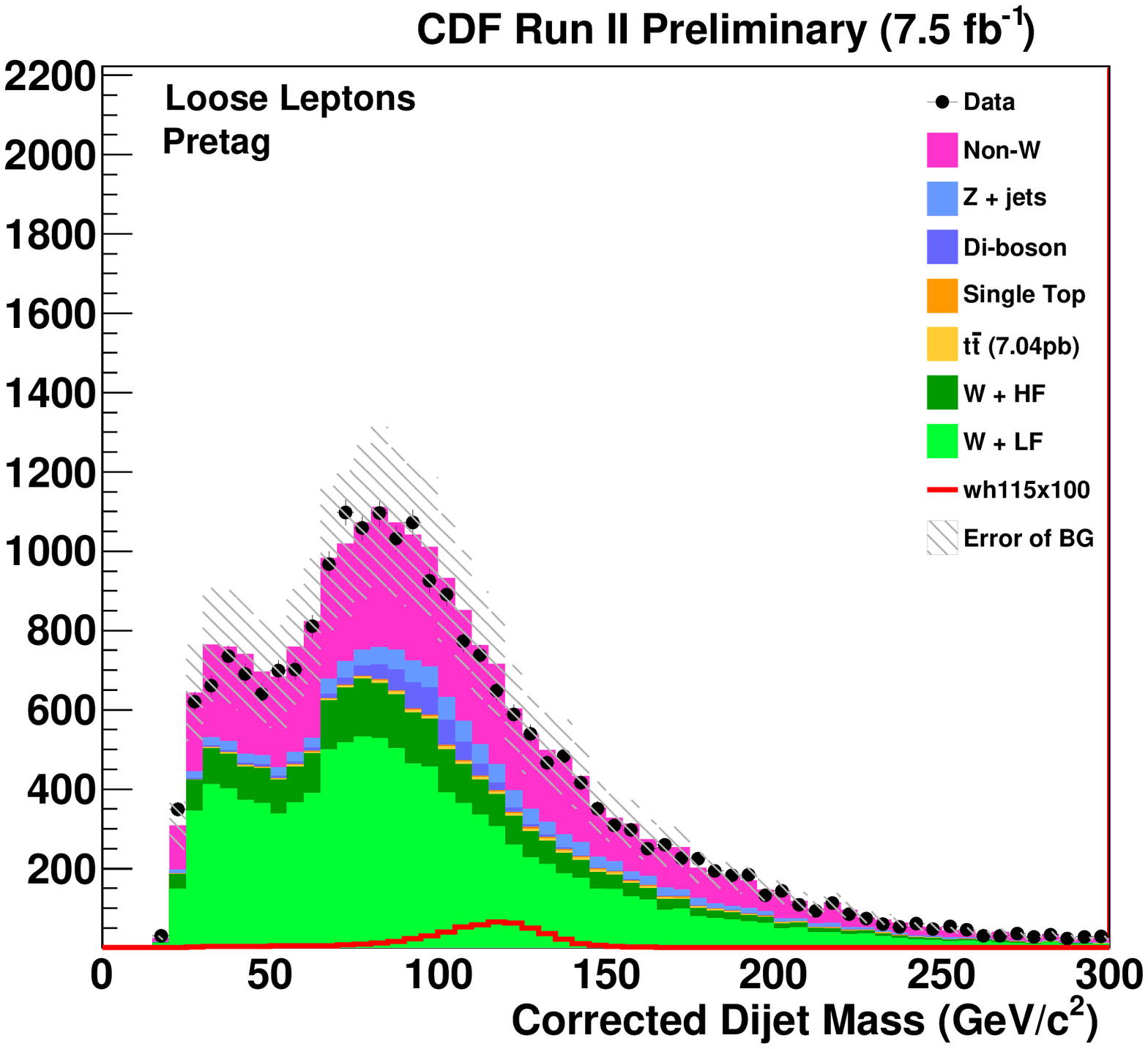}

\caption{(a) The transverse momentum $p_{T}$ distribution of $W$ boson candidates in the
central lepton selected sample of the CDF experiment and (b) the transverse energy imbalance (MET) in the two-jet selected sample of the D0 experiment. (c) The dijet mass distribution in two-jet 
selected events from the D0 search, after application of a new multivariate discriminant to suppress 
multijet background and (d) the dijet mass distribution after correcting for dijet mass resolution 
effects in loose lepton selected events by the CDF experiment. Both the CDF and D0 
collaborations select four initial search samples and make use of loose lepton selection criteria. 
The candidate $W\rightarrow e \nu $ and $W\rightarrow \mu \nu $ events are shown combined in the figures. } \label{preselection}
\begin{picture}(0,0)(0,0)
\put (-245,455){\scriptsize {\bf (a)}}
\put (-40,455){\scriptsize {\bf (b)}}
\put (30,275){\scriptsize {\bf (d)}}
\put (-245,275){\scriptsize {\bf (c)}}
\end{picture}
\end{figure*}

Candidate $W\rightarrow e \nu $ and $W\rightarrow \mu \nu $ events are selected by requiring a 
single isolated lepton together with an associated imbalance in transverse energy in the events. 
Jets are reconstructed using iterative cone algorithms which make use of midpoints as additional seeds. Figure \ref{preselection}(a) shows the $p_{T}$ distribution of $W$ candidates selected in events 
with two large-$p_{T}$ centrally reconstructed leptons by the CDF Collaboration.\cite{cdf} The search is based on a total integrated luminosity of $\cal{L}$$=7.5 ~\rm fb^{-1}$. Prior to the application
of $b$-tagging, a total of four samples are selected using different leptonic angular and 
reconstruction criteria, each requiring two reconstructed large-$p_{T}$ jets. The recently added 
fourth search sample is based on loose lepton selection criteria.

Figure \ref{preselection}(b) shows the transverse energy imbalance in the $W$ candidate events 
in $\cal{L}$$=7.5 ~\rm fb^{-1}$ of data selected by the D0 Collaboration.\cite{d0} Four samples are 
selected prior to the application of $b$-tagging, by requiring either 
two or three reconstructed jets for each leptonic channel. The three-jet selected
sample is included to allow for additional gluon radiation in the collision hard subprocess.
Loose lepton selection criteria are used for both the electron and muon 
channel searches.

The background contributions in each sample are normalized according to their theoretical predictions and/or modeled in separately selected control samples. The {\tt PYTHIA} \cite{pythia}, {\tt ALPGEN} \cite{mlm}
or {\tt COMPHEP} \cite{comphep,comphep1} event generators are used, with {\tt ALPGEN} and 
{\tt COMPHEP} interfaced to {\tt PYTHIA} to account for subsequent hadronization according to
the MLM factorization (``matching'') scheme.\cite{mlm} Both the CDF and D0 selections include 
additional smaller sensitivities to $W\rightarrow \tau \nu$ decays, in which the $\tau$ subsequently decays into an electron or a muon. In addition, candidate
$ZH\rightarrow \ell \ell b\bar{b}$ events in which one decay lepton passes and one fails the isolated 
lepton selection criteria used in $ZH\rightarrow \ell \ell$ searches are also selected.

Background from multijet (QCD) production processes is studied separately using the data. 
Figure \ref{preselection}(c) shows the dijet invariant mass distribution of the D0 Collaboration after 
application of a multivariate discriminant technique to further improve suppression of multijet 
background. The CDF experiment employs a similar approach based on the super vector machine 
technique. Figure \ref{preselection}(d) shows the dijet invariant mass distribution for the loose 
selected lepton sample by the CDF experiment after applying recent improvements to improve dijet mass
resolution. The improvements, which are from 15\% to 11\%, have a direct impact in the procedure to 
set upper limits and are described in detail in \cite{dijet}. 
A good description of the data by the sum of the expected search backgrounds is obtained in all cases.

\begin{figure*}[ht]
\centering
\includegraphics[width=93mm]{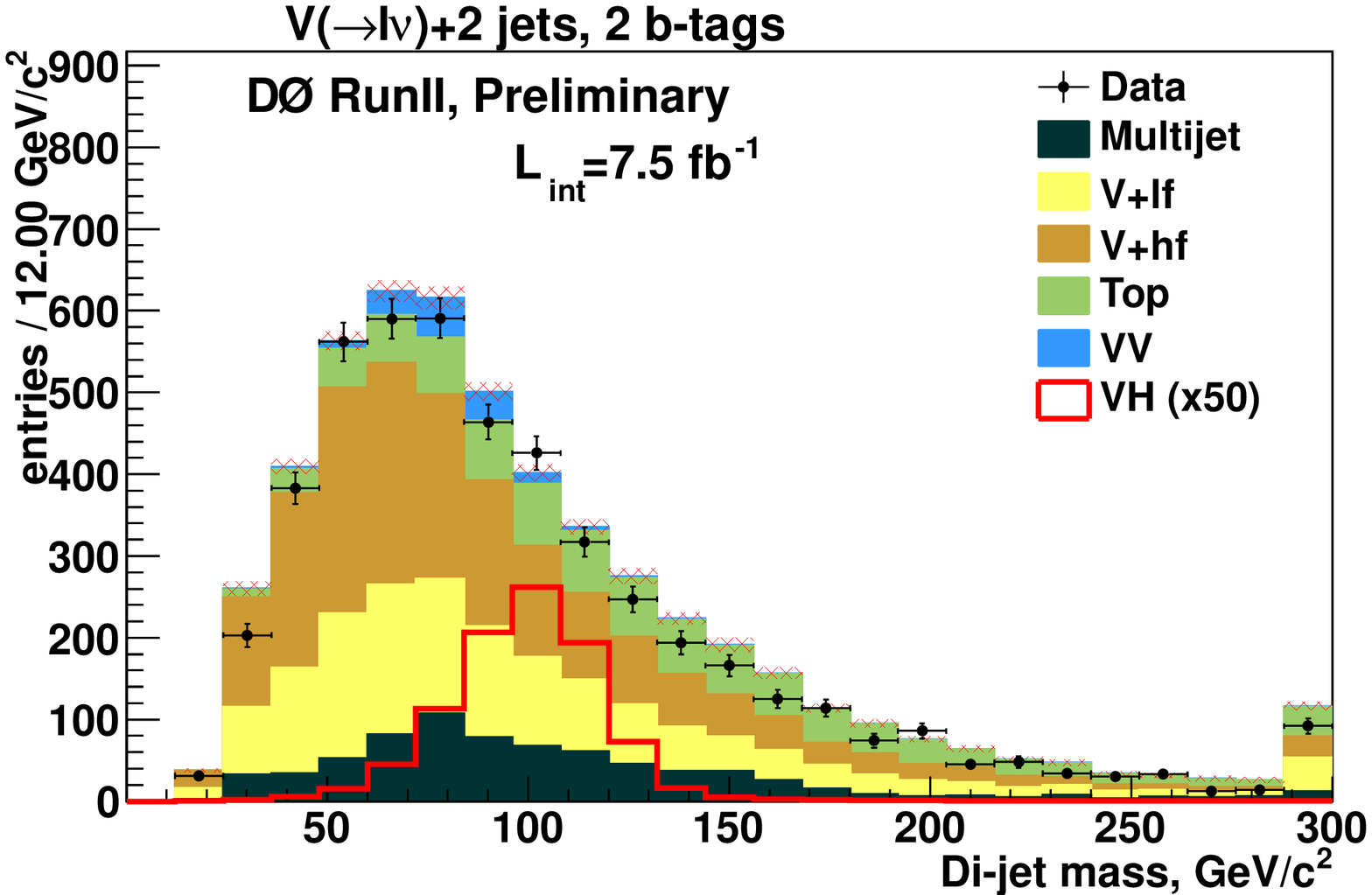}
\includegraphics[width=93mm]{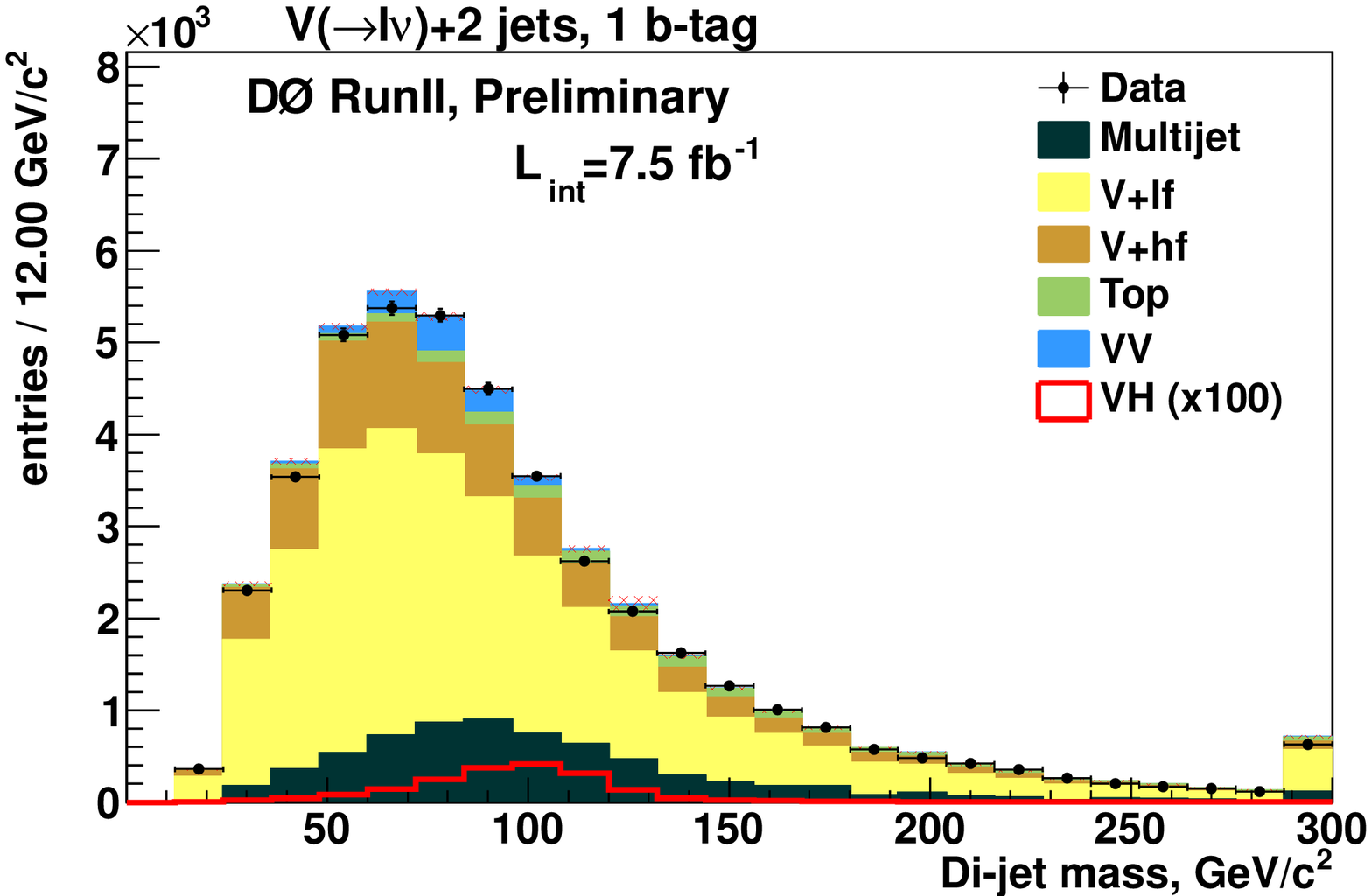}
\caption{Data samples divided into statistically independent samples using $b$-tagging. (a) The 
dijet invariant mass distribution in two-jet, two $b$-tagged jets events from the D0 collaboration 
and (b) the dijet mass distribution in two-jet events containing a single $b$-tagged jet.  
The expected signal contribution for a Higgs mass
$M_{H}=115 ~\rm GeV$ is shown scaled in the figures and the electron and muon leptonic search 
samples are combined.} \label{btagging}
\begin{picture}(0,0)(0,0)
\put (-140,415){\scriptsize {\bf (a)}}
\put (-140,235){\scriptsize {\bf (b)}}
\end{picture}
\end{figure*}

\section{Application of $b$-tagging}

The events are further subdivided into statistically independent
(orthogonal) samples through the application of $b$-tagging. The obtained
samples are of different sensitivities due to the different background contributions
which remain in each sample.

The D0 Collaboration uses a Neural Network approach to tag $b$-jets which is based on a total 
of seven input discriminating variables.\cite{d0btag} Two samples are selected from each of the 
previously selected two-jet and three-jet samples. The first sample is selected by requiring
two jets to be consistent with having been initiated by $b$-quarks. Events that
fail the first condition are used to select a second sample requiring a single $b$-tagged jet.  
Figure \ref{btagging}(a) shows the D0 dijet invariant mass distribution 
in the two-jet, two $b$-tagged sample (the 
electron and muon leptonic channels are shown combined in the figure).
Figure \ref{btagging}(b) shows the dijet invariant mass distribution in two-jet events that fail 
the two $b$-tag requirement and contain a single $b$-tagged jet. 
The expected signal contribution for a 
Higgs mass $M_{H}=115 ~\rm GeV$ is shown scaled by factors of 50 and 100 in each figure, respectively.

The CDF Collaboration tag $b$-jets using a combination of three separate techniques; (1) a neural
network algorithm (NN), (2) secondary vertex tagging (ST) and (3) requiring that the probability
for a jet to have originated from the primary vertex (JP) to be small. Four independent samples are
selected, three of which require two $b$-tagged jets and one of which requires a single $b$-tagged 
jet. The two-jet samples are selected using different combinations (ST-ST, ST-JP and ST-NN)
of the $b$-tagging requirements.

\begin{figure*}[ht]
\centering
\includegraphics[width=80mm]{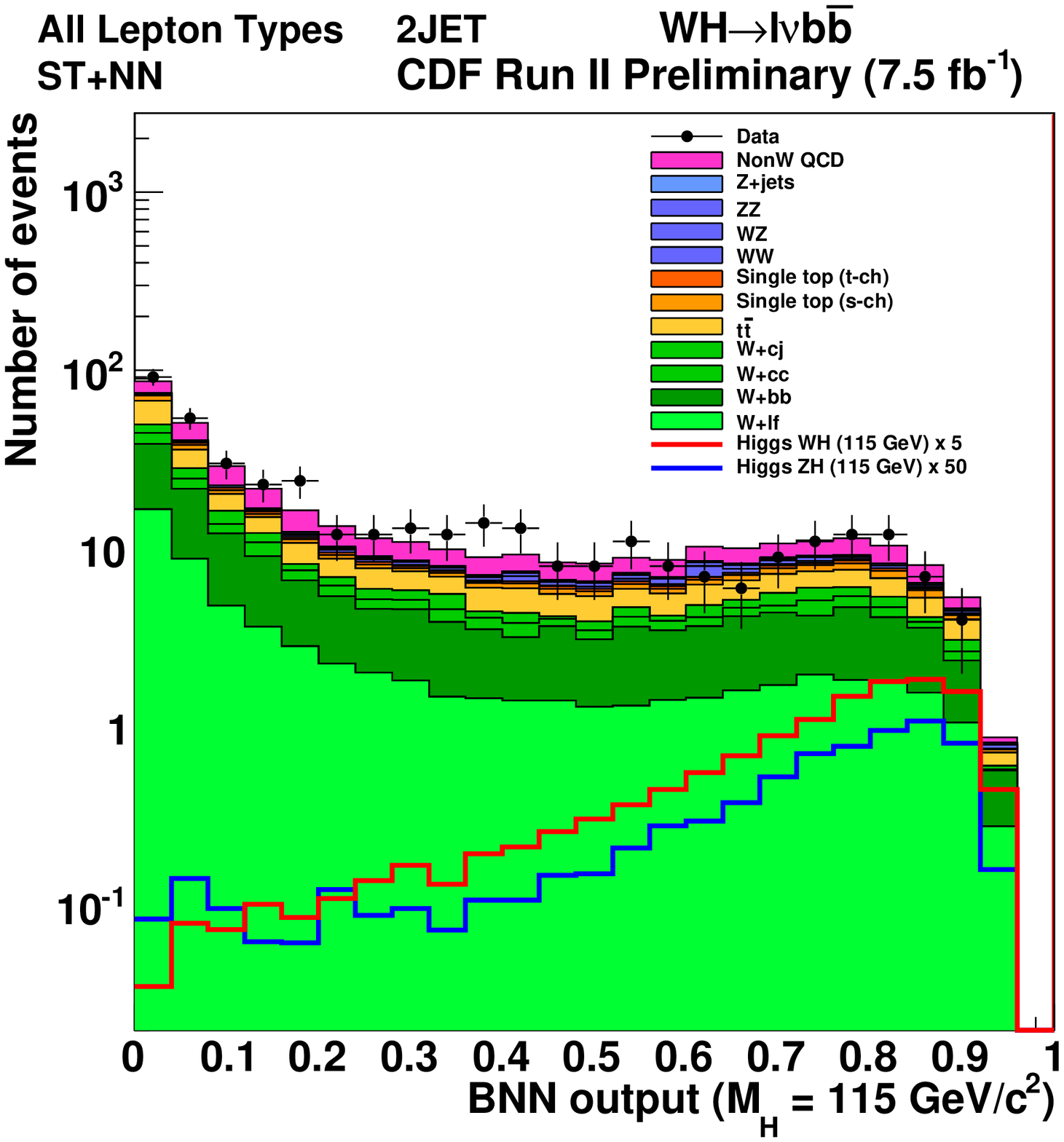}
\includegraphics[width=80mm]{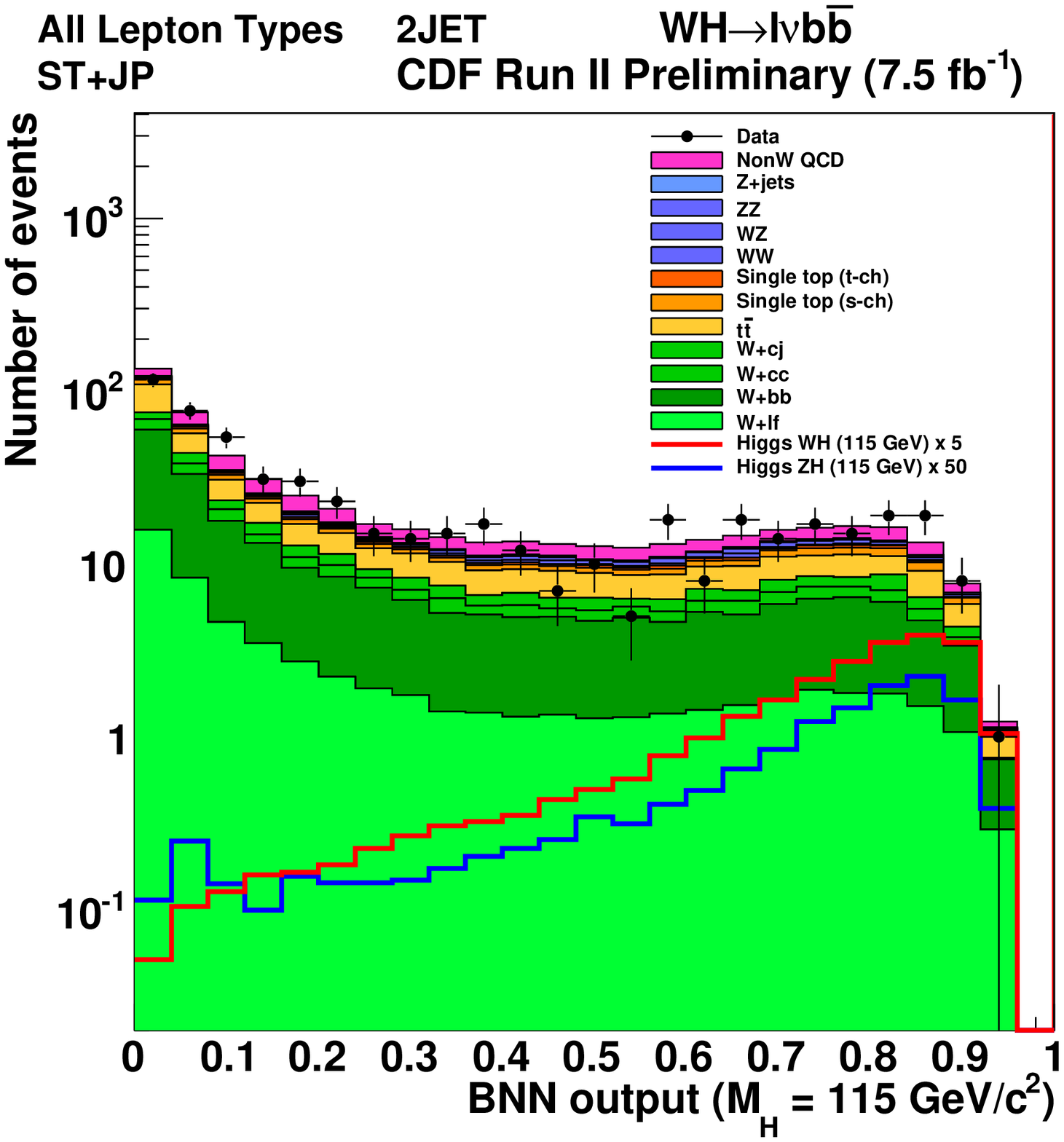}
\caption{Multivariate techniques used to suppress remaining backgrounds in the $b$-tagged samples. 
The output distribution of the Bayesian Neural Network (BNN) for (a) the ST+NN (secondary vertex + Neural Net) and (b) the ST+JP (secondary vertex + jet probability) $b$-tagged jet samples of the CDF Collaboration. The expected $WH$ signal contribution is shown scaled by a factor 5 in each figure and 
the leptonic search criteria have been combined.} \label{mva}
\begin{picture}(0,0)(0,0)
\put (-235,285){\scriptsize {\bf (a)}}
\put (-10,285){\scriptsize {\bf (b)}}

\end{picture}
\end{figure*}

\section{Multivariate Discrimination}

The remaining backgrounds in each $b$-tagged sample are suppressed through 
the application of multivariate techniques. The CDF Collaboration use a 
Bayesian Neural Network (BNN) approach which is based on eight discriminating input variables. 
The eight input event variables are studied and optimized separately for each of the orthogonally 
selected $b$-tagged samples.  Figures \ref{mva}(a) and (b) show the output distributions 
obtained after applying the BNN to the CDF ST-NN and ST-JP $b$-tagged samples, respectively. 
In each distribution, the $q\bar{q}\rightarrow WH$ signal peaks at large discriminant output values, 
whereas the sample backgrounds are shifted towards small discriminant output values. 
The four leptonic selections are shown combined in each figure and the expected
$WH$ signal contribution is shown scaled by a factor 5.

The D0 Collaboration use a Boosted Decision Tree (BDT) approach, which is applied 
multiple times to each $b$-tagged sample to generate a random forest. A total of 13 
input variables are used for each decision tree. These are randomly assigned from a total 
of 20 event input variables, which are studied and optimized 
in separate studies. Since the approach is able to discriminate against multiple background sources, 
the same 20 input event variables are used as discriminants for each of the orthogonally 
selected $b$-tagged samples.

\begin{figure*}[ht]
\centering
\includegraphics[width=155mm]{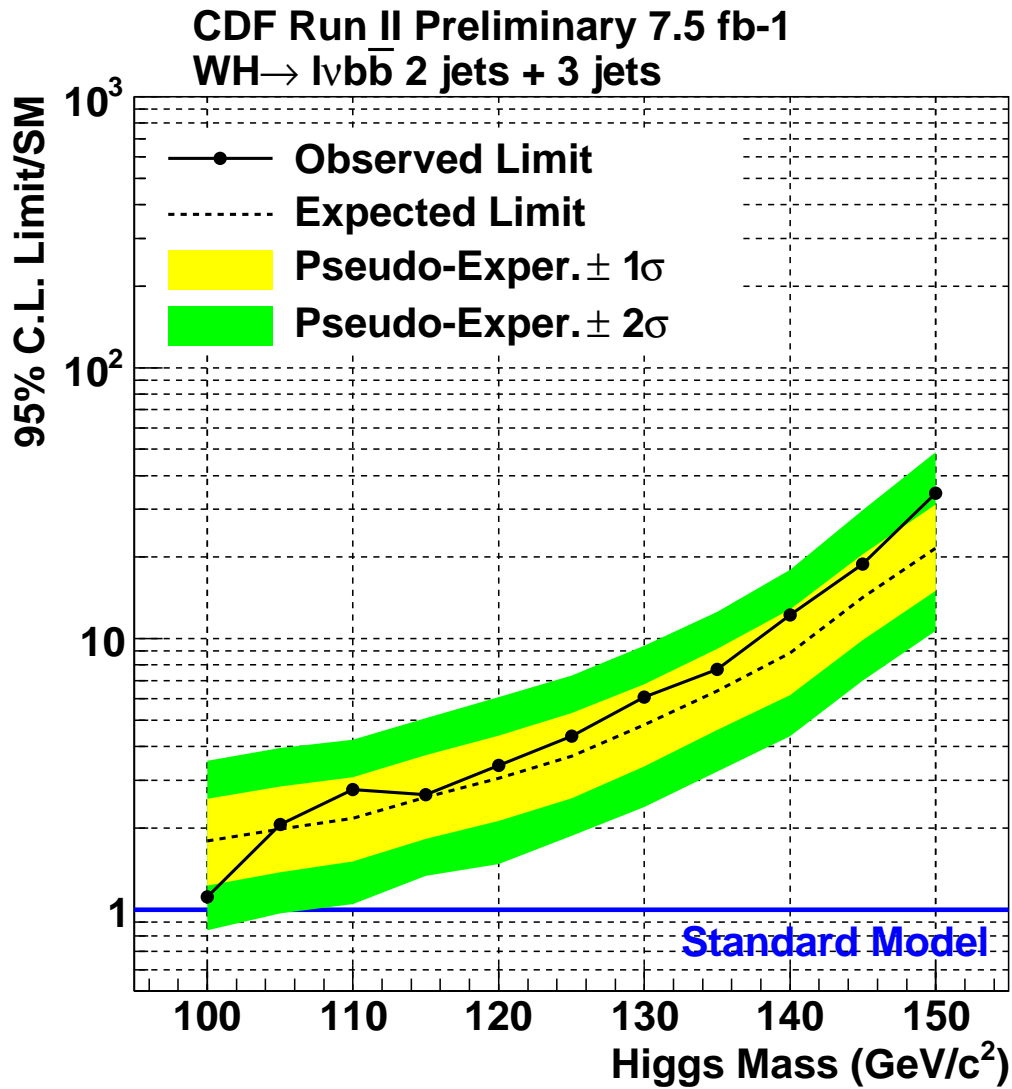}
\caption{CDF observed (solid line) and expected (dashed line) upper limits for the $WH$ cross 
section times branching ratio in units of the SM prediction. The two-jet search 
is combined with the independent CDF three-jet search of \cite{CDFlimit}. The bands incorporate the effect of the systematic and Poisson statistical uncertainties} \label{cdf_limits}
\end{figure*}

\section{Cross Section Upper Limits}

Upper limits are derived for product of the $WH$ production cross section and
branching ratios and reported in units of the SM prediction. The output 
multivariate distributions described in the previous section are used as 
discriminating inputs in the limit derivation procedure.

Figure \ref{cdf_limits} shows the observed and expected upper limits
of the CDF collaboration search. The results are shown combined with an 
independent search based on a three-jet event selection by the CDF
Collaboration which uses a matrix method approach for backgrounds.\cite{CDFlimit}
The observed limits are derived for 11 discrete values of Higgs mass $M_{H}$.
The expected limits are shown by the dashed line and the bands incorporate
the effect of the systematic and Poisson statistical uncertainties. 
The observed (expected) upper limits for a Higgs mass 
$M_{H}=115 ~\rm GeV$ are 2.65 (2.6), respectively, and represent a 17\% 
improvement in total sensitivity.

\begin{figure*}[h]
\centering
\includegraphics[width=155mm]{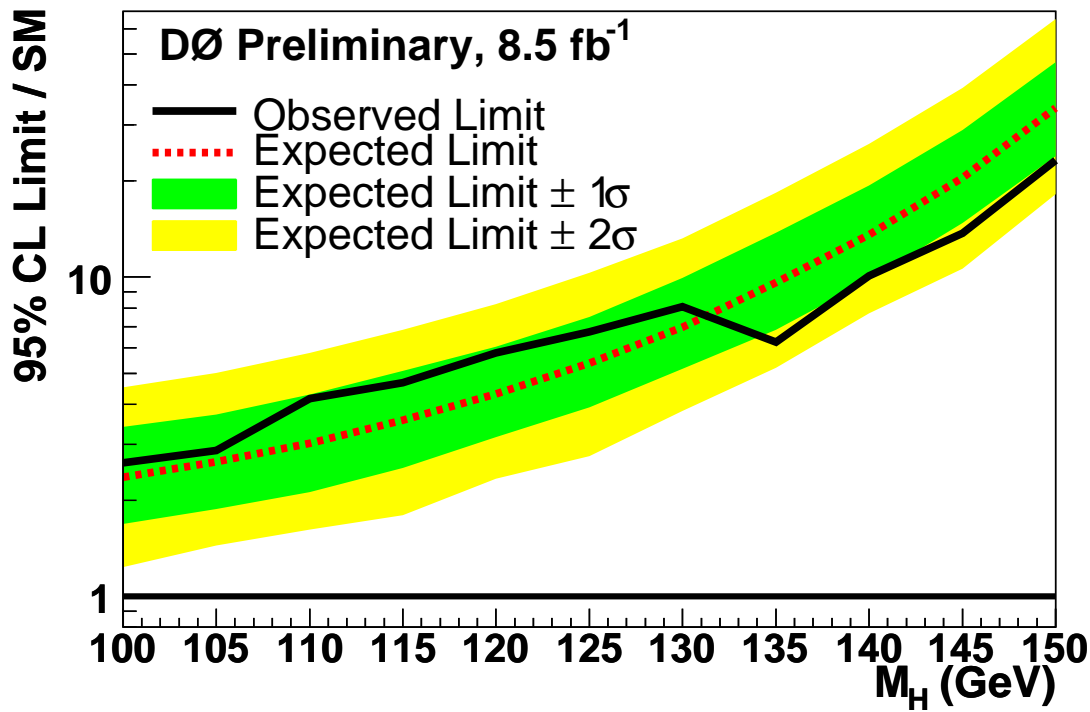}
\caption{D0 observed (solid line) and expected (dashed line) upper limits for the $WH$ cross section 
times branching ratio in units of the SM prediction. The search 
is combined with the $1 ~\rm fb^{-1}$ D0 result previously reported in \cite{D0limit}. 
The bands incorporate the effect of the systematic and Poisson statistical uncertainties} \label{d0_limits}
\end{figure*}

The observed and expected upper limits of the D0 Collaboration search 
are shown in Fig. \ref{d0_limits}. The results are combined with a 
previously published D0 result, which is based on $1~\rm fb^{-1}$ of analyzed data.\cite{D0limit}
The bands incorporate the effect of the systematic and Poisson statistical uncertainties
and the observed limits are again derived for 11 discrete values of Higgs mass.
The observed (expected) upper limits for a Higgs mass of $M_{H}=115 ~\rm GeV$
are 4.6 (3.5), respectively, and the results represent an 11\% improvement beyond those
expected from the increase in analyzed luminosity alone.

\section{Summary}

Searches for $WH\rightarrow \ell \nu b \bar{b}$ final states are particularly sensitive to 
SM Higgs production at Tevatron $p\bar{p}$ Collider energies. The
selection of event samples containing an isolated lepton, an imbalance in transverse
energy in the events, and either one or two reconstructed jets consistent with having
evolved from a $b$-quark, provides statistically independent data samples to search for
$q\bar{q}\rightarrow WH$ candidates. The observed (expected) limits for a Higgs mass of $M_{H}=115 ~\rm GeV$ are  2.65 (2.6) and 4.6 (3.5) for the CDF and D0 searches, respectively, and
the results, which are based on large luminosity data samples collected by the CDF and D0 collaborations, continue to gain in sensitivity.

%\begin{acknowledgments}
%\end{acknowledgments}

\bigskip % extra skip inserted
%\begin{thebibliography}{9}   % Use for  1-9  references

\end{document}